\documentclass[]{aa}

\usepackage{epsfig}
\usepackage{textcomp}
\usepackage{amsfonts}
\usepackage{amssymb}
\usepackage{natbib}

\usepackage{color}
\usepackage{graphicx}
\usepackage{psfrag}

\newcommand{\refe}{}
\newcommand{\refee}{}

\begin{document}

\title{Impact of inclination on quasi-periodic oscillations \\from spiral structures}
\author{P. Varniere\inst{1}
\and F. H. Vincent\inst{2,3}}

\institute{AstroParticule \& Cosmologie (APC), UMR 7164, 
Universit\'e Paris Diderot, 10 rue Alice Domon et Leonie Duquet, 75205 Paris Cedex 13, France.
varniere@apc.univ-paris7.fr
\and     Observatoire de Paris / LESIA, 5, place Jules Janssen, 92195 Meudon Cedex, France.
\and  Nicolaus Copernicus Astronomical Center, ul. Bartycka 18, PL-00-716 Warszawa, Poland.}

\date{Received  /
Accepted }
 
\abstract
{Quasi-periodic oscillations (QPOs) are a common feature of the power spectrums of X-ray binaries. Currently it is not  possible
to unambiguously  differentiate the large number of proposed models  to explain these phenomena through existing observations.}
{We investigate the observable predictions of a simple model that generates \refe{flux modulation}: a spiral
instability rotating in a thin accretion disk. This model is motivated by  the accretion ejection instability (AEI)
model for low-frequency QPOs (LFQPOs). We are particularly interested in the inclination dependence
of the observables that are associated with this model.}
{We develop a simple analytical model of an accretion disk, which features a spiral instability.
\refe{The disk} is assumed to emit blackbody radiation, which is ray-traced to a distant observer.
We compute pulse profiles and power spectra as observed from infinity.}
{We show that the amplitude of the modulation associated with the spiral rotation is a strong
function of inclination and frequency. The pulse profile is quasi-sinusoidal only at low inclination (face-on source).
As a consequence, a higher-inclination geometry leads to a stronger and more diverse harmonic signature in the power spectrum.}
{We  present how the amplitude depends on the inclination when the flux modulation comes from a spiral in the disk.
We   also include new observables that could potentially differentiate between models, such as 
the pulse profile and  the harmonic content of  the power spectra of high-inclination sources that exhibit LFQPOs.
These might be important observables to explore with existing and new instruments.
}

\titlerunning{Observables to differentiate QPO models}
\maketitle

\section{Introduction}  

Quasi-periodic oscillations (QPOs) are a  common feature of  the power spectrum of black-hole
X-ray binaries~\citep[see e.g. the reviews by][and references therein]{remillard06,done07}. 
\refe{Early on,  QPOs were observed outside of the X-ray band, for example in the optical~\citep{Motch83} and recently in the infrared~\citep{Kalamkar15}.}
They are characterized by a Lorentzian component in the
power spectrum of the source, distinguished from broader noisy peaks by
a condition on the coherence parameter ($Q = \nu_{\mathrm{LF}} / \mathrm{FWHM} \gtrsim 2$,
with many cases at $Q>10$).
Depending on their frequencies they are   either called a high-frequency QPO ($\sim 40$ to $500$~Hz) 
or low-frequency QPO for frequencies within approximately $[0.1 - 30]$~Hz.
LFQPOs are seen when the power-law component of the spectrum is strong (such as the hard, hard/intermediate states). 
This suggests that these oscillations are not only linked to the thermal disk but also to a non-thermal component
which could be either a hot inner flow~\citep{done07} or some kind of a hot corona above the disk.
Different types of LFQPOs have been introduced based mainly on their characteristics.
In this article we are focusing on the most frequent one, namely the type-C LFQPO, which has the interesting property of having a strongly varying 
frequency $\nu_{\mathrm{LF}}$ that seems to be correlated with  the disk flux. This may be a hint that LFQPOs have an origin in the disk while also 
being reprocessed in a corona or hot inner flow.

While we have a lot of observations, no consensus on the origin of the modulation has yet been achieved. There are several classes of models, depending
on what causes the modulation. 
Among all the families of models, one originated from~\citet{stella98} which advocated that the Lense-Thirring precession of particles around a rotating black hole could produce LFQPOs.  This family was further extended by~\citet{schnittman06} and more recently~\citet{ingram09}  and~\citet{nixon14}.
Another family is based on different kinds of instabilities occuring in either the disk or the corona. 
Among those \citet{tagger99} looked at a spiral instability driven by magnetic stresses (the {accretion-ejection instability}, AEI).
\citet{chakrabarti00} propose that QPOs are due to radiation emitted by oscillating shocks in  the disk while~\citet{titarchuk04} consider a transition layer that comptonizes the outer disk.
In yet another approach,~\citet{cabanac10} consider a magneto-acoustic wave propagating in the 
corona while~\citet{oneill11} present MHD simulations of blackhole accretion flow exhibiting a 
dynamo cycle with a characteristic time that coincides with LFQPOs.
\refe{While several models have links  with emission outside  the X-ray band, for example in the case of the AEI~\citep{VT02}, the most developed 
model at the moment is the one  by~\citet{Veledina13,Veledina15},  which explains this multi-wavelength phenomena by elaborating on the precessing hot flow model of~\citet{ingram09}.
It is worth noting that their curve of rms as a function of the inclination of the system is quite different from the one presented here, and observation might be able to 
distinguish between them.}

        Despite the amount of ongoing modeling, up to now QPO models have  focused mainly on explaining the frequency observed in the 
        power spectrum (PDS).  
     While this is essential, it is only the first observable we had access too and
     all of the models fulfilled this requirement since they were created to answer it. It is therefore interesting to  look  for other observables that might help to differentiate between models.
     Some of those we already have access to through observation, such as the amplitude, but also some we do not have yet complete access to, such as the pulse profile.

In this article we are interested in revisiting the AEI model and expand on a \refe{previous Newtonian approach}~\citep{varniere05} to include 
a wider range of frequencies and general-relativistic effects. In particular, we want to see how the observables are affected by the general relativistic effects in higher-inclination systems.
Already with this simple model, we aim at studying the evolution of the amplitude
of the spiral modulation as a function of the QPO frequency
and to do this for high- and low-inclination systems. 
We are also interested in how this affects the  shape of the pulse profile and hence the harmonic content of its power spectrum.
Section~\ref{sec:model} presents  our simple model of a disk featuring a spiral instability. Section~\ref{sec:rms}
shows the evolution of the amplitude with inclination and QPO frequency. Section~\ref{sec:pulse}
analyzes the pulse profile for various inclinations and discusses the corresponding
power spectra.

\section{\refe{The accretion-ejection instability as a model for LFQPO}}

\refe{
In a nutshell, the accretion-ejection instability is a global spiral instability which occurs in the inner region of a fully magnetized (close to equipartition) accretion disk \citep{tagger99}. 
The energy is extracted from the disk (hence allowing accretion) through the spiral wave and then stored in a Rossby vortex, which is located at the corotation radius of the spiral 
(the point where the spiral and accretion disk rotate at the same velocity), which is typically a few times the inner radius of the disk. 
 In the presence of a low-density corona, the Rossby vortex will twist the
 footpoint of the magnetic field line. This causes an Alfven wave to be emitted toward the corona, therefore linking accretion and ejection \citep{VT02} 
 which, in turn, has the unique consequence of linking what is happening in the thermal disk with what is happening at higher energy in the corona. 
This way the modulation in the disk will be transmitted to the corona. }
\newline

\refe{
Here we will summarize salient points and observational tests of the AEI as the origin of the LFQPO. We focus on the X-ray data which represent the bulk of the
observations:
\begin{description}
   \item[\tt -] the rotation frequency of the spiral is the orbital frequency at the corotation radius, $r_c$,
                which is predicted to be a few tenths of the orbital frequency at the inner edge of the disk. 
                This frequency is consistent with the LFQPO frequency \citep{tagger99}.
   \item[\tt -] as the position of the inner edge of the disk evolves during the outburst, the rotation frequency of
   the spiral  also changes, which can be directly compared with observation \citep{varniere02}.
   \item[\tt -] by including general relativity through the existence of a last stable orbit and orbital velocity profile, the AEI explains the 
                observed turnover in the correlation between the color radius   (= inner disk radius, as determined by the spectral fits) and the LFQPO 
                frequency \citep{rodriguez02,mikles09}.
   \item[\tt -] in the Newtonian approximation, the AEI is able to create a thermal flux modulation \citep{varniere05} in the range of the observed one.
   \item[\tt -] once the AEI is established, 2D MHD simulations show a nearly steady rotation pattern, and  is thus able to account for 
              persistent LFQPOs
   \item[\tt -] the AEI transfers energy and angular momentum toward the corona by Alfven waves, thus providing a 
              supply of Poynting flux that may produce the compact jet often observed in the low-hard state  \citep{VT02}.
   \end{description}
        Because of these characteristics, the model based on the assimilation of the AEI with the origin of the LFQPO has since been expanded, first as a scenario for the
         $\beta$ class of  GRS~1915+105 \citep{T04}, then as a way of classifying blackhole states \citep{V11} and, more recently, a possible explanation
         for the different types of LFQPO \citep{V12}. 
}

\section{Amplitude of the modulation for a spiral model}
\label{sec:model}
   
     Rather than looking at the formation of the spiral instability  in a disk, we decided to focus on its consequence
     on the emission. 
     Consequently, rather than taking full MHD simulations of the AEI  in the different conditions we wish to explore, 
     we decided to create a simple, analytical model that mimics the temperature profile of the  AEI to test the different parameters more cleanly.   
     Indeed, in a full fluid simulation, changing one parameter in the initial condition can have repercussions on several observable parameters and, therefore,  
     it is harder to study the different effects separately. 
     We consider a geometrically thin accretion disk,  introduce a spiral feature in this disk at a higher temperature,  and consider the emission of blackbody radiation.  
     \refe{ Also, through the AEI, energy and angular momentum is sent toward the corona~\citep{VT02}, but the exact amount depends on many parameters 
that are not easily constrained by observation. Here we take a simpler approach; rather than adding multiple unconstrained parameters, we
decide to stick to a very simple, analytic model of the thermal emission in the disk for which we have more constraints. 
This means that we are  able to predict the light curve of 
LFQPOs in the lower energy part of the spectrum ($1$keV), leaving a more in-depth study of the LFQPO rms-spectrum as a function of energy for a more complex, 
simulation-based, model. 
}
     
    We have therefore taken a simple, analytic model of a disk that features a spiral structure, which is able  to reproduce the main aspects of the AEI model as 
    seen in simulations \citep[see for example][]{V12}, and see how the flux is thus modulated.

\subsection{Parametrization of the disk temperature profile}

We consider a geometrically thin disk surrounding a Schwarzschild blackhole of mass $M$.
The disk extends from a varying inner radius $r_{\mathrm{in}}$ to a fixed outer radius $r_{\mathrm{out}} =500\,M$. 
\refee{To keep a simple structure, we choose to have the temperature profile $T_0(r)\propto r^{-\eta}$. In agreement with the thin disk blackbody model,
we took $\eta=0.75$ as the equilibrium temperature profile.}
This profile is \refee{then} fixed by choosing the temperature at the innermost stable circular orbit (ISCO), labeled $T_{\mathrm{ISCO}}$.

This equilibrium disk is assumed to give rise to a spiral instability. We model this phenomena in a very simple
way, just describing the hotter spiral structure rotating in the equilibrium disk, in agreement with numerical simulations of the AEI~\citep[see for example][]{V12}. 
The temperature of the disk with this added spiral feature reads
  \begin{eqnarray}
   \label{eq:T}
 T(t,r,\varphi) &=& T_0(r) \\ \nonumber
 &&\times\left[  1 + \gamma \left(\frac{r_c}{r}\right)^\beta\mathrm{exp}\left(-\frac{1}{2}\left(\frac{r-r_s(t,\varphi)}{\delta \,r_c}\right)^2\right)\right]^2 , \\ \nonumber
  \end{eqnarray}
where the perturbation term between brackets describes the spiral pattern.
The parameter $\gamma$ encodes the temperature contrast between the spiral and the surrounding disk.
The quantity $r_c$ is the corotation radius of the spiral.
The spiral temperature is thus following a power-law decrease when moving away from $r=r_c$, with an exponent $\beta$. This ensures
that the spiral will fade away into the disk after a few turns. In the latter Gaussian term, $r_s$ is a shape function 
encoding the spiral feature. It ensures that the spiral's width is a factor $\delta$ times the corotation radius.
The shape function reads
      \begin{equation}
      r_s(t,\varphi) = r_c \,\mathrm{exp}\left(\alpha (\varphi-\Omega(r_c)t)\right) , 
      \end{equation}
where $\alpha$ is the spiral opening angle and $\Omega(r_c)$ is the Keplerian
frequency at $r_c$. Ultimately it is the rotation frequency of the spiral and the frequency at which the flux is modulated.

  \subsection{Model parameters}

        The spiral parameters are chosen to (1) be in agreement with numerical simulations of the AEI and 
        (2) produce reasonable values of amplitude. The present study is not devoted to the detail of the
        dependency of observables on the model parameters \citep[see][for a parameter study in the Newtonian approximation]{varniere05}. 
        Instead, we  fix the parameters that represent the spiral and then study, in a frozen framework, how the inclination impacts the amplification, and this for 
        a sample of modulation frequencies representative of the type-C LFQPO.
        
        In this respect, the temperature at the ISCO is taken to be $T_{\mathrm{ISCO}} = 10^7$~K, thus emitting blackbody radiation mainly around $1$~keV
        and the disk is geometrically thin with an aspect ratio of $H/r = 0.01$.
        The power-law exponent encoding the temperature decrease away from $r=r_c$ is set to $\beta=0.5$.
        This ensures that the spiral will be negligible within a few turns. 
        The opening angle is chosen to be $\alpha=0.1,$ in agreement with numerical simulations~\citep{V11,V12}.
        Then we decided to use and freeze the parameters of the spiral so that a QPO of  about $10$~Hz had an amplitude of  \refee{about $10\%$} at an
        inclination of $70^\circ$. This gives \refee{$\gamma=2.2$} and $\delta=0.2,$ \refee{which are conservative compared} to the maximum 
        reached in numerical simulations \citep{V12}.  These same parameters gave an \refee{amplitude of about $3\%$} at an inclination of $20^\circ$. 
        
        In  the rest of the paper, only the inclination at which the system is observed and the position of the inner edge of the disk will be varied.
        This  produces a large set of QPO frequencies.
        Indeed, as shown by~\citet{tagger99}, the corotation radius of the spiral is a few times the inner radius of the disk.
        Here we chose to keep $r_{c}/r_{\mathrm{in}}$ fixed at a typical value of $2$, so that, as the inner radius varies, so
        the corotation radius.
        We  consider an inner edge of the disk varying in the range \refee{ $[1.5\,r_{\mathrm{ISCO}}, 15\,r_{\mathrm{ISCO}}]$ which \refe{creates
        modulation frequencies in the range $[1\,\mathrm{Hz}, 42\,\mathrm{Hz} ]$ for $M=10\,M_{\odot}$. }}
        \refe{We do not compute very-low-frequency QPOs (below $1~$Hz) because, in our model, they demand both
a very large disk (the corotation radius for this kind of frequency is far from the last stable orbit) and a very fine grid resolution, which translates into overly long computing time. 
We also consider  one simulation at \refee{42~Hz}, thus above the maximum observed LFQPO frequency,  to capture the tendency of the rms amplitude towards the highest frequency LFQPOs.}

\subsection{Emission from the disk and ray-tracing}
     
     The whole disk is assumed to simply emit as a blackbody at the temperature $T(t,r,\varphi)$. So the specific intensity emitted at some position in the disk is
     \begin{equation}
     I_\nu^{\mathrm{em}} = B_\nu(\nu^{\mathrm{em}},T) , 
     \end{equation}
     where the superscript em refers to the emitter's frame, i.e. a frame corotating (at the local Keplerian frequency) with the disk. This emitted intensity is then
     transformed to the distant observer's frame using the constancy along geodesics  of $I_\nu / \nu^3$. Thus
     \begin{equation}
     I_\nu^{\mathrm{obs}} = g^3 I_\nu^{\mathrm{em}} , 
     \end{equation}
     where $g=\nu^{\mathrm{obs}}/\nu^{\mathrm{em}}$ is the redshift factor. This redshift factor is, in particular, responsible for the so-called beaming
     effect, which makes the observed specific intensity stronger when the emitter  travels towards the observer and fainter in the opposite case.
     \refe{In the following we  use $h \nu^{\mathrm{obs}} = 1~$keV.}
     
     To compute maps of specific intensity $I_\nu^{\mathrm{obs}}$, we use the open-source general relativistic ray-tracing code GYOTO~\citep{Vin11} 
  into which we added the parametrized disk profile that was defined in the previous section. Null geodesics are
  integrated in the Schwarzschild metric, backwards in time from a distant observer at some inclination with respect
  to the disk. Inclination is equal to the angle between the observer's line of sight and the normal to the black hole's
  equatorial plane. From such maps of specific intensity, the light curve (flux as a function of time) is derived by summing all pixels weighted
  by the element of solid angle, which is subtended by each pixel.

\begin{figure}[htbp]
 \centering
\includegraphics[width=0.4\textwidth,clip]{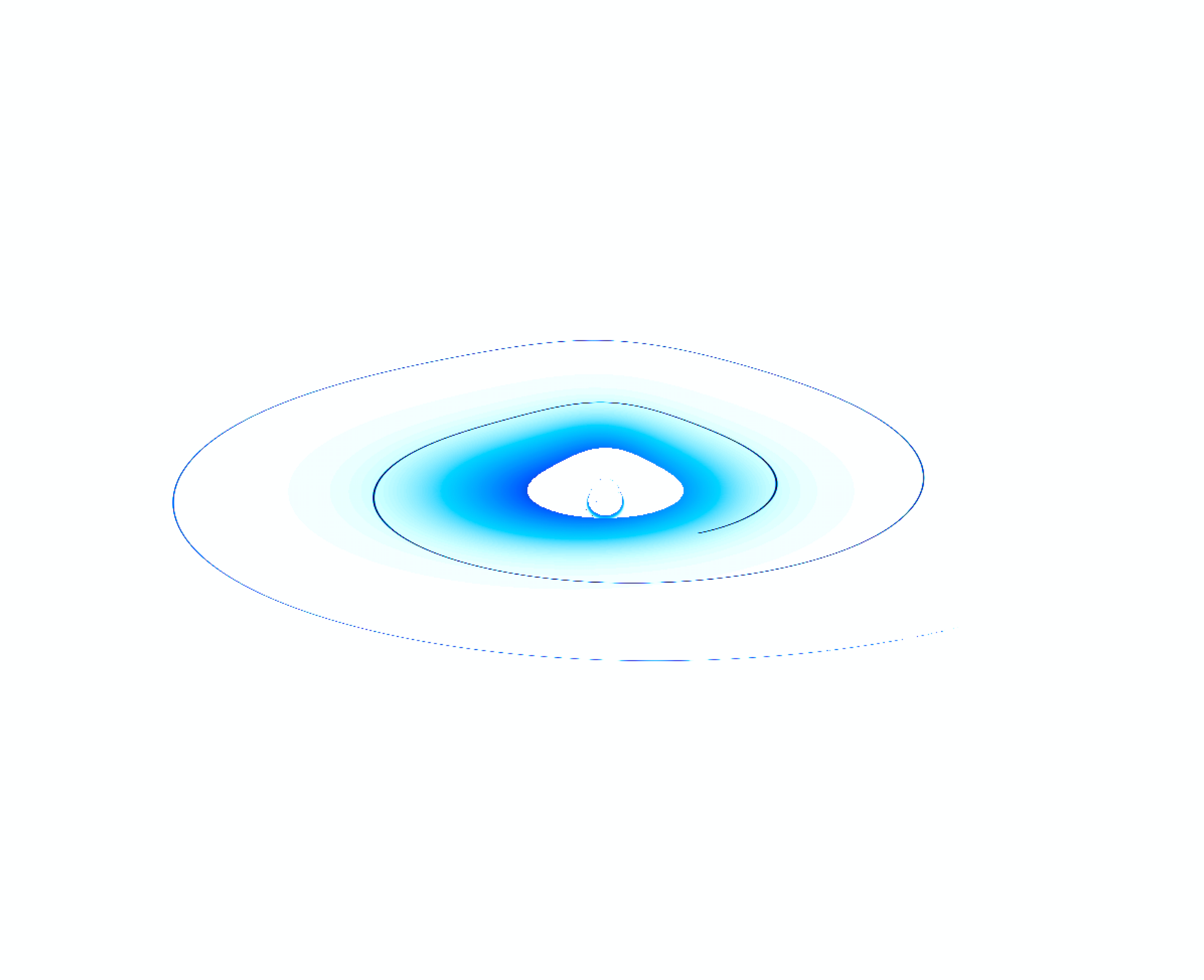}
 \caption{\refee{Logscale view of the} inner region of an accretion disk, having a hotter $m=1$ spiral wave with
 a corotation radius of \refee{$r_c=10\,r_{\mathrm{ISCO}}$}, seen at an inclination of $i=70^{\circ}$.}
  \label{fig:spiral}
\end{figure}

     Figure~\ref{fig:spiral} shows, \refee{in logscale}, a spiral with a corotation radius located at \refee{$r_c=10\,r_{\mathrm{ISCO}}$.}
     It shows the spiral arm fading into the background disk following the power-law decrease of the amplitude as the spiral
     expands. 
     Before it reaches the end of the disk, the spiral has fully faded. Most of the emission actually comes from the first two turns of the spiral, close to $r=r_c$, and we only consider
     the full spiral  for consistency.

\section{Impact of frequency and inclination on the amplitude}
\label{sec:rms}

        We can now take advantage of our analytical model and test how the amplitude of the modulation that comes from the rotating spiral
         behaves with respect to the position of the corotation radius, namely the frequency and the inclination of the system. While it seems easy to 
        obtain observational data in the first case (amplitude versus $\nu_{QPO}$), the second case (amplitude versus inclination of the system) seems less attainable. 
        Here we need to clarify that, even this first case is not as easily compared to observation as it seems. Indeed,  we are trying to look at the
        impact of one parameter at a time, therefore no temporal evolution is taken into account. We are not "monitoring"\ the evolution of the instability as
        the inner edge of the disk moves but, instead, looking at {\it the same spiral} in different locations, meaning all the other parameters are frozen while we take
        snapshots with different inner edges of the disk. \\

        To be able to compare the different lightcurves on the same plot  we  renormalize them  with the time being $t/T$ where $T$ is the period of each modulation,
        and the flux being $f/f_{\mathrm{mean}}$, where $f_{\mathrm{mean}}$ is the mean flux of each particular curve. 
        This  allows for an easier comparison of the shape of each lightcurve independantly of its total flux and period.
        For all models, we define the amplitude as
 \begin{equation}
     \mathrm{amp} = \frac{f_{\mathrm{max}}-f_{\mathrm{min}}}{f_{\mathrm{max}}} ,
\end{equation}
     where $f_{\mathrm{min}}$ and $f_{\mathrm{max}}$ are the extremal values of flux.

 \subsection{Origin of the modulation}

        In the case of a spiral, or any non-axisymetrical structure orbiting in the disk, one would get a modulation as 
        a combination of two effects. First,  as the spiral orbits around the disk, its observed intensity is  modulated by the beaming
       effect: this will be stronger on the approaching side and fainter on the receding side. 
       In addition, the projected  emitting area on the observer's sky varies with time as the spiral rotates, which  also creates a flux modulation.
       From this, it appears that the most important parameter of the spiral, as far as amplitude is concerned, 
      is its temperature contrast with the disk (encoded in the $\gamma$ parameter). The higher the contrast with the disk, 
      the greater the  amplitude of the modulation. 

\begin{figure}[htbp]
 \centering
\includegraphics[width=0.5\textwidth,clip]{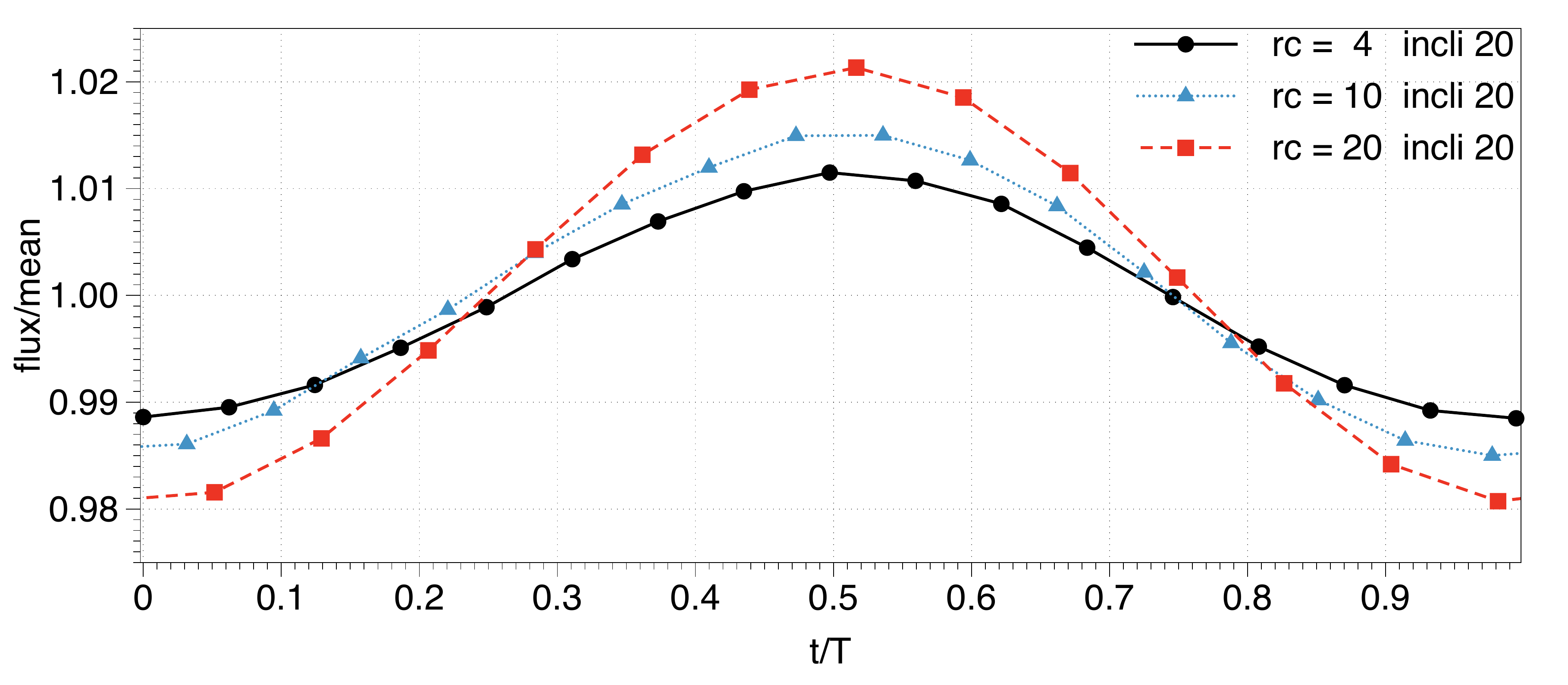}  
\caption{Renormalized light curves obtained for the same spiral with three values of the inner edge of the disk and associated corotation
radii.  Here the system is viewed at  an inclination $i=20^\circ$ \refee{and the black dots represent $r_c=4 r_{ISCO}$, blue triangles $r_c=10 r_{ISCO}$ and red squares $r_c=20 r_{ISCO}$.}
  }
\label{fig:LC_spiral_rc}
\end{figure}

        In Figure~\ref{fig:LC_spiral_rc} we show the renormalized lightcurves for three positions of the inner edge (and consequently
        of the spiral's corotation radius) for a disk 
        seen under an inclination of $20^\circ$ (close to face-on). Several facts already appear on this plot. While all of the spirals
        have the same amplitude, $\gamma$, the amplitude increases as the \refe{corotation radius} is further away in the disk.

         This can easily be  understood as follows. Given the temperature profile in 
        Eq.~\ref{eq:T}, it is straightforward to obtain that the ratio between the maximum temperature
        along the spiral (i.e. at the corotation radius $r=r_c$) and the temperature at the inner edge
        is $T_{\mathrm{spi}} / T_{\mathrm{in}} =  (1+\gamma)^2/2$ with \refee{$\gamma=2.2$} in our case. 
        It is then easy to compute
        the corresponding ratio of blackbody emission $B_\nu(T_{\mathrm{spi}}) / B_\nu(T_{\mathrm{in}})$ 
        as a function of the value of the inner radius. This ratio is a strongly increasing function of
        the inner radius. As a consequence, the spiral dominates more and more the inner regions of
        the disk as the inner radius increases. Thus, {\it at similar spiral parameters, the amplitude increases with a receding disk}. 
        It is important to note that this is valid for similar spiral parameters and that, for widely varying spiral parameters,
        the amplitude behavior with respect to the inner edge of the disk could be reversed.
        \refe{On the same note, if we relax the constraint on the corotation 
        $r_c = 2 r_{in}$\footnote{In the AEI framework, this would be the equivalent of saying that the disk physical paramaters, such as the density and magnetic field, are changing.},  
        we can see for the same $r_c$ and spiral parameters, the modulated part of the flux is constant while, changing the position of the inner edge of the disk changes 
        the unmodulated part of the flux, hence impacting the total rms amplitude of the modulation. }

        Another interesting point is that, at low inclination, the modulation closely resembles a sine function. It takes a Fourier
        decomposition to see the presence \refee{of small amplitude harmonics and is often limited to the first few harmonics}  (more about this in Section \ref{sec:pulse}).
         Those also become stronger similarly to the amplitude when the inner edge of the disk is further away from the ISCO.

\subsection{Impact of QPO frequency on the amplitude}

        \refe{Following the observation from Fig.\ref{fig:LC_spiral_rc}. that the same spiral  has a stronger amplitude as it is 
        further away in the disk,
        we can plot the evolution of the amplitude of the modulation} created by the same-parameter spirals
        as a function of the couple ($r_{\mathrm{in}}$, $r_c=2\,r_{\mathrm{in}}$). With the parameters we fixed,   
        we get an approximate \refee{$3$\%} amplitude at a frequency of $10$~Hz for the low inclination ($i=20^\circ$) case. 

\begin{figure}[htbp]
 \centering
\includegraphics[width=0.5\textwidth,clip]{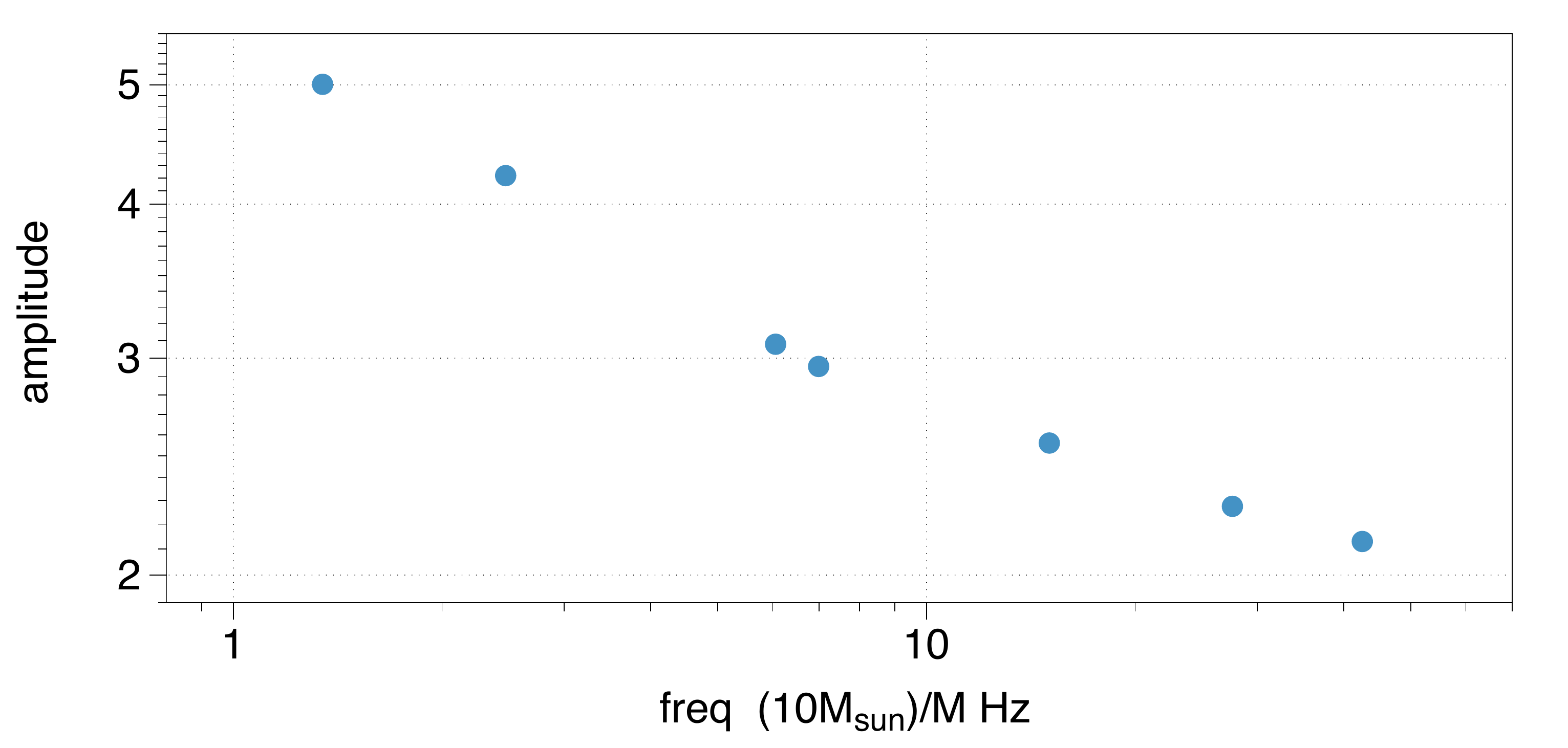}
\caption{Evolution of the amplitude of the modulation as function of the QPO frequency, meaning $\Omega(r_c = 2\,r_{\mathrm{in}})$,  for an inclination $i=20^\circ$.
}
\label{fig:rms_vs_freq_incli20}
\end{figure}

        As we can see in Fig.~\ref{fig:rms_vs_freq_incli20}, we follow a wide range of frequencies and see a steady increase 
        in the amplitude as the frequency gets lower. 
\refe{This is in part linked to  the fact that we are keeping the inner egde of the disk and the corotation radius in a locked ratio of $r_c=2r_{in}$. Indeed,
 as the frequency increases it means that the inner edge of the disk gets closer to the last stable orbit, hence gets hotter and emits more in the observation band. 
 As a consequence the modulation amplitude over the total emission is lower. 
}

        Here we need to be cautious, since this plot is done for the same spiral parameters but looking at different
        frequencies, meaning that it does not represent an evolution of the spiral responsible for the modulation. 
        In the case of a temporal evolution, such as the one would get in a full MHD simulation and/or a full outburst, the spiral would also change 
        (e.g. its amplitude)
        hence it would influence the precise shape of this plot. Nevertheless, it shows a strong propensity of the amplitude to decrease with 
        frequency.

\subsection{Amplitude dependence on inclination}

     One parameter that can change a lot, both 
     the beaming effect and the projected emitting surface (and thus, the modulation) 
     is the inclination at which the system is observed. Using observation alone this is not something that can easily be studied.      
     However, \citet{motta15} show one of the first statistical studies demonstrating that higher-inclination systems tend to have a higher 
      amplitude for the QPO than is the case for lower-inclination systems. 
     It is therefore interesting to determine how the  amplitude in an almost edge-on system (inclination $70^\circ$) and 
     an almost face-on system (inclination $20^\circ$) would compare for {\it exactly the same spiral parameters}.      
     For this reason we  focus on the ratio between amplitudes at different inclinations and not their actual value. 
        
\begin{figure}[htbp]
 \centering
\includegraphics[width=0.5\textwidth,clip]{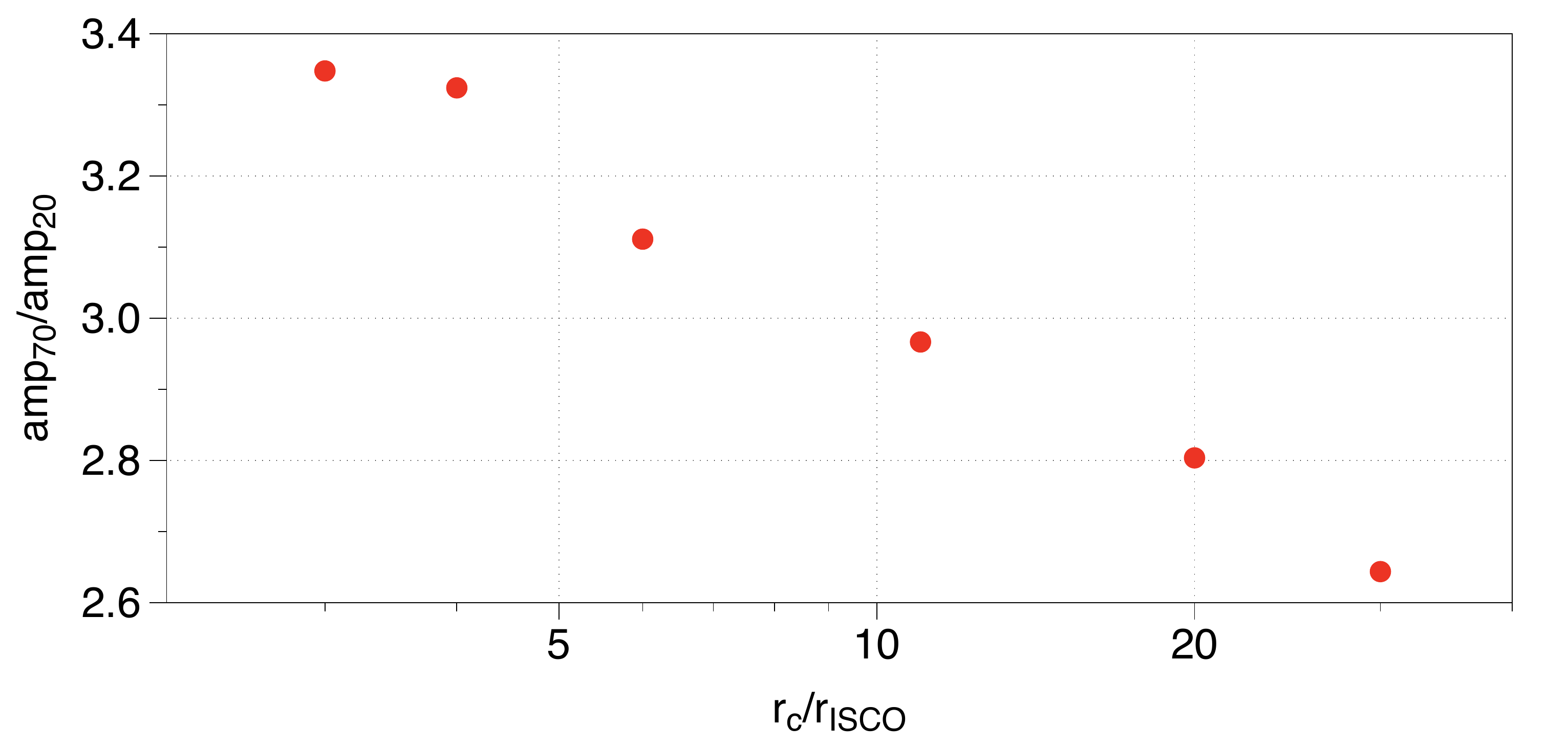}
 \caption{ Evolution of the ratio of the amplitude at inclination $70^{\circ}$ over the amplitude at inclination $20^\circ$ as a function of the position of the structure given in units of the  ISCO radius.}       
\label{fig:ratio_rms_incli}
\end{figure}

        Figure~\ref{fig:ratio_rms_incli} shows the evolution of the ratio of the amplitude at inclination $70^{\circ}$ over the amplitude at inclination $20^\circ$  
        as a function of the position of the structure in the disk, namely $r_c$, 
        which is related to the frequency through the mass of the central object.
        Indeed, as in observation, one cannot observe the same system at different inclinations, it seems easier to produce a plot that could be  
        scaled to the \refee{different masses} of the objects.
        While high inclination system always have a higher amplitude than low inclination system for the same physical
parameters, we see
        than the ratio is dependant on the frequency of the modulation, linked to the position of the co-rotation radius.
        We see that the ratio decreases from \refee{$
         \sim 3.5$} for small $r_c$, which translates to a higher frequency, to \refee{$
         \sim 2.5$} for  larger 
        $r_c$, meaning a lower frequency. 
        Indeed, this is related to the fact that the flux modulation is, in particular, due to the beaming factor.
        The flux will be boosted when the hottest parts of the spiral
        are on the approaching side of the disk, while the flux will be deboosted when these hottest parts are on the receding side.
        This boosting is a strong function of inclination: it will be bigger for close to edge-on views because the
        emitter will then be traveling almost exactly towards the observer.
        Furthermore, as we get far from the blackhole, the effect of the beaming will decrease, and   will thus be less important for
        lower QPO frequencies (higher $r_c$).
        
        Figure~\ref{fig:ratio_rms_incli}  is impossible to compare directly with the results of  \citet{motta15}  since these authors average several sources, hence different 
        blackhole masses, which in turn mean the same QPO frequencies do not correspond to the same distance in the disk,
        and to different inclinations and different times of their outburst evolution.
        On the contrary, here we  look at the amplitude dependence on inclination for exactly the same system 
        at different QPO frequencies, which is clearly not what happens in an outburst, since the source evolves dynamically. 
        Nevertheless, it is already interesting to see that the same system, under the same conditions, does indeed have a different amplitude,
        depending on the inclination of the system, and that this difference  evolves, depending on the position of the spiral in the disk.

\section{Case of the pulse profile in high-inclination systems}
\label{sec:pulse}

        While the previous section focused on the  amplitude of the QPO, there is more to QPOs than just a frequency and an amplitude.     Indeed, here we have access to the exact pulse profile. This allows us to study the more direct impact of inclination on the lightcurve.
     
\subsection{Shape of the pulse}

\begin{figure}[htbp]
 \centering
\includegraphics[width=0.5\textwidth,clip]{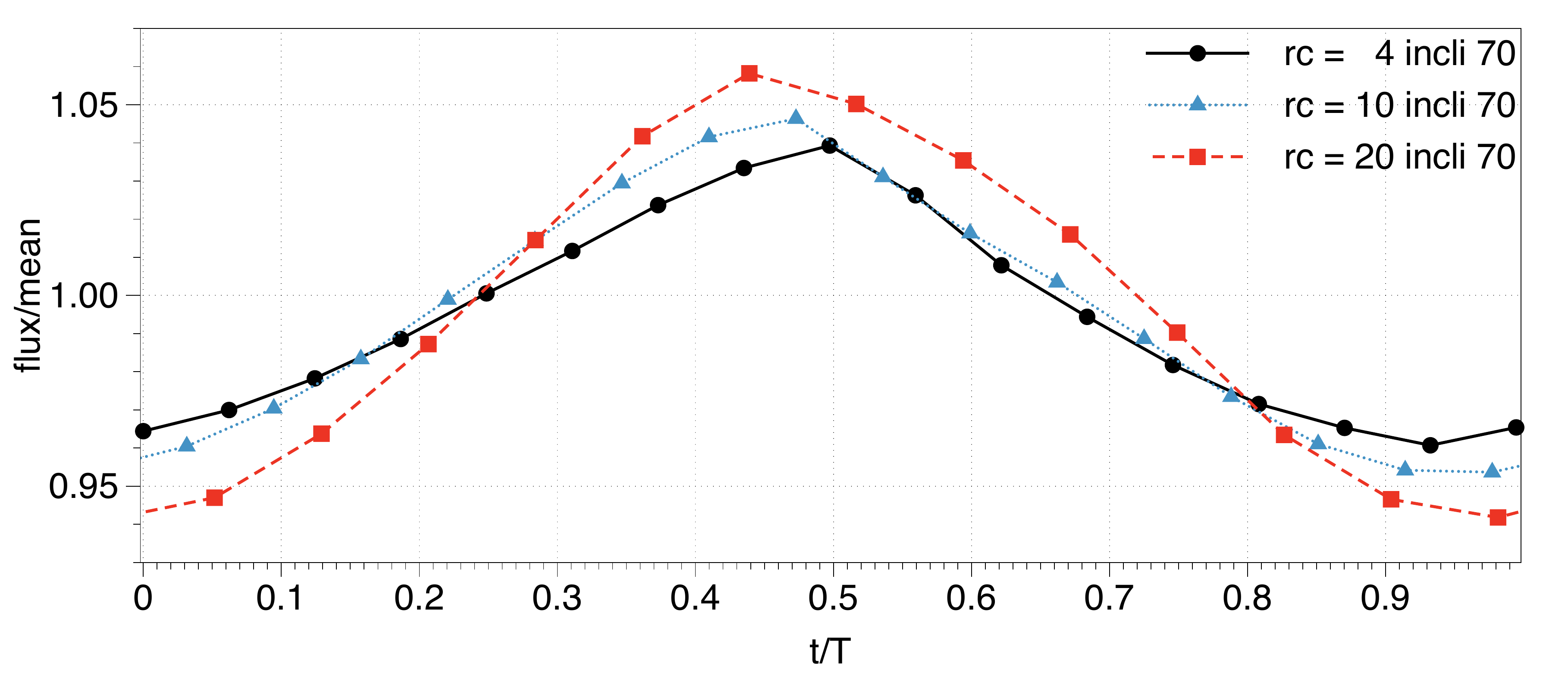}
 \caption{Same as Fig.~\ref{fig:LC_spiral_rc} for an inclination of $70^\circ$.}
  \label{fig:LC_spiral}
\end{figure}

        In Fig.~\ref{fig:LC_spiral}, we show one period of three pulse profiles for three different locations of the spiral, as seen from an inclination
        of $70^\circ$. We have chosen the same position/frequency as in Fig.~\ref{fig:LC_spiral_rc} so that it is easier to compare  both cases.
        The main difference with the $20^\circ$  inclination case is that, at high inclination, the detailed shape of the pulse is clearly not sinusoidal.
        This implies that any Fourier decomposition will contain more than one frequency, even when the initial signal in our model only has  one.
        This effect comes from  Doppler boosting and the varying apparent area of the spiral on the observer's sky, 
        and it gets stronger with inclination. 
        We  explore how this should be visible on the PDS in the next section .

        However, it is worth mentioning another possible way of detecting this change in the shape of the pulse profile, even without having access
        to its details. Indeed,  in Fig.~\ref{fig:LC_spiral} we see that for high-inclination systems,
        \refee{the lightcurves are not symmetrical with respect to the mean values.}
        While this is \refee{only a few \% and} hardly detectable all the time, we may find some high-inclination systems that have a strong QPO around a few Hz in a 
        steady enough observation in which we can try to assess the time spent above and below the mean value. 
                Perhaps this would be worth investigating in, for example, the Plateau state of GRS $1915$+$105$? While the resolution needed to look
        at it on a pulse timescale is beyond the capacity of past and present satellites, it would be an interesting new observable to differentiate
        between models for future missions.

\subsection{Impact on the PDS}

        As seen in the previous section, the shape of the pulse profile departs more and more from a sinusoid as the inclination of the system increases from
        almost face-on to almost edge-on. When looking at it in the Fourier space, this departure  translates into more harmonics that might be more easily detectable than
        a difference in the pulse profile. 

        To compute the power density spectrum, we extended our lightcurve to $20$ orbits and added a $1/f$ noise.
        Also, as we are interested in comparing the extent of the harmonic structure with respect to inclination, we  re-normalized the amplitude of the
        fundamental peak in each case to $1$. This allows us to compare only the relative strength of the different peaks.
        
        \refe{In the case of a \refee{QPO frequency of about $2.5$~Hz (hence a corotation radius of $20\,r_{\mathrm{ISCO}}$), we already have a first harmonic
        present even at the inclination of $20^\circ$, while the pulse profile looks sinusoidal (see Fig.\ref{fig:LC_spiral_rc}.).} }
        In the case of a close to edge-on view, we get a much richer and stronger harmonic content. Indeed, \refee{we have not one, as in the close to face-on view,
        but several additional\refee{ and stronger} peaks,} while the simulation we ran had one mode present (one-arm spiral). 
        This behavior of richer harmonic content for high-inclination systems does not seem to depend on the frequency of the modulation, \refe{though it would be easier to 
        detect for higher amplitude fundamentals, hence smaller frequencies.}
         
        However, care must be taken when interpreting this result since instabilities that give rise to spiral waves in a disk tend to have more than one
        mode and therefore the original signal could be more complex than the case with only one mode that is presented here. The different modes
        are in a close to harmonic relationship and  could therefore lead to harmonic peaks in the PDS. Nevertheless, these instabilities also
        tend to have such a degree of competition between modes (more as a transition from one to another) and it should not be difficult to detect this.

\section{Conclusion}

In this paper we looked at a simplified version of the spiral formed by  accretion ejection instability and studied
the lightcurves that are observable from this type of system.
Our main goal was to study how the  amplitude of the observed lightcurve  depends  not only on the frequency of the QPO, but also on  its
inclination, with respect to the observer. 

Our model predicts that, for a similar spiral structure, at a given frequency, the amplitude is strongly dependent on inclination and increases at high inclination.
Using parameters in agreement with both numerical simulations and observations, our simple model is able to produce amplitudes 
that differ by an amount of \refee{$\sim 2.5 - 3.5$} for high-inclination ($70^\circ$) with respect to low-inclination ($20^\circ$) sources.
Thus it is in agreement with the recent study of~\citet{motta15},
which shows higher amplitudes from high-inclination sources, although it is not possible to directly compare the actual amplitude ratio 
predicted by our simulations  with observations, since we do not know if the observed sources were  in a state that is similar to our model.

Another interesting point raised by our ability to produce detailed lightcurves is how the pulse profile of these QPO changes with
frequency and inclination. It becomes less and less sinusoidal as the source inclination
increases. This translates into a richer harmonic content in the power spectrum. 
While there is no statistically significant proof that high-inclination sources have a significantly higher harmonic content than low inclination sources,
 it is coherent with the data gathered in~\citet{motta15} (private communication). This would be worth exploring in more detail, especially
 if other QPO models have distinct predictions.

\begin{acknowledgements}
PV acknowledges financial support from the 
UnivEarthS Labex program at Sorbonne Paris Cit\'e (ANR-10-LABX- 0023 and ANR-11-IDEX-0005-02). FHV acknowledges financial support by  
the Polish NCN grant 2013/09/B/ST9/00060.
Computing was partly done using the Division Informatique de l'Observatoire (DIO) HPC facilities from Observatoire de Paris (http: //dio.obspm.fr/Calcul/) 
and at the FACe (Francois Arago Centre) in Paris.
\end{acknowledgements}

\bibliography{Spiral}

\begin{thebibliography}{25}
\expandafter\ifx\csname natexlab\endcsname\relax\def\natexlab#1{#1}\fi

\bibitem[{{Cabanac} {et~al.}(2010){Cabanac}, {Henri}, {Petrucci}, {Malzac},
  {Ferreira}, \& {Belloni}}]{cabanac10}
{Cabanac}, C., {Henri}, G., {Petrucci}, P.-O., {et~al.} 2010, \mnras, 404, 738

\bibitem[{{Chakrabarti} \& {Manickam}(2000)}]{chakrabarti00}
{Chakrabarti}, S.~K. \& {Manickam}, S.~G. 2000, \apjl, 531, L41

\bibitem[{{Done} {et~al.}(2007){Done}, {Gierli{\'n}ski}, \& {Kubota}}]{done07}
{Done}, C., {Gierli{\'n}ski}, M., \& {Kubota}, A. 2007, \aapr, 15, 1

\bibitem[{{Ingram} {et~al.}(2009){Ingram}, {Done}, \& {Fragile}}]{ingram09}
{Ingram}, A., {Done}, C., \& {Fragile}, P.~C. 2009, \mnras, 397, L101

\bibitem[{{Kalamkar} {et~al.}(2015){Kalamkar}, {Casella}, {Uttley}, {O'Brien},
  {Russell}, {Maccarone}, {van der Klis}, \& {Vincentelli}}]{Kalamkar15}
{Kalamkar}, M., {Casella}, P., {Uttley}, P., {et~al.} 2015, ArXiv e-prints

\bibitem[{{Mikles} {et~al.}(2009){Mikles}, {Varniere}, {Eikenberry},
  {Rodriguez}, \& {Rothstein}}]{mikles09}
{Mikles}, V.~J., {Varniere}, P., {Eikenberry}, S.~S., {Rodriguez}, J., \&
  {Rothstein}, D. 2009, \apjl, 694, L132

\bibitem[{{Motch} {et~al.}(1983){Motch}, {Ricketts}, {Page}, {Ilovaisky}, \&
  {Chevalier}}]{Motch83}
{Motch}, C., {Ricketts}, M.~J., {Page}, C.~G., {Ilovaisky}, S.~A., \&
  {Chevalier}, C. 1983, \aap, 119, 171

\bibitem[{{Motta} {et~al.}(2015){Motta}, {Casella}, {Henze},
  {Mu{\~n}oz-Darias}, {Sanna}, {Fender}, \& {Belloni}}]{motta15}
{Motta}, S.~E., {Casella}, P., {Henze}, M., {et~al.} 2015, \mnras, 447, 2059

\bibitem[{{Nixon} \& {Salvesen}(2014)}]{nixon14}
{Nixon}, C. \& {Salvesen}, G. 2014, \mnras, 437, 3994

\bibitem[{{O'Neill} {et~al.}(2011){O'Neill}, {Reynolds}, {Miller}, \&
  {Sorathia}}]{oneill11}
{O'Neill}, S.~M., {Reynolds}, C.~S., {Miller}, M.~C., \& {Sorathia}, K.~A.
  2011, \apj, 736, 107

\bibitem[{{Remillard} \& {McClintock}(2006)}]{remillard06}
{Remillard}, R.~A. \& {McClintock}, J.~E. 2006, \araa, 44, 49

\bibitem[{{Rodriguez} {et~al.}(2002){Rodriguez}, {Varni{\`e}re}, {Tagger}, \&
  {Durouchoux}}]{rodriguez02}
{Rodriguez}, J., {Varni{\`e}re}, P., {Tagger}, M., \& {Durouchoux}, P. 2002,
  \aap, 387, 487

\bibitem[{{Schnittman} {et~al.}(2006){Schnittman}, {Homan}, \&
  {Miller}}]{schnittman06}
{Schnittman}, J.~D., {Homan}, J., \& {Miller}, J.~M. 2006, \apj, 642, 420

\bibitem[{{Stella} \& {Vietri}(1998)}]{stella98}
{Stella}, L. \& {Vietri}, M. 1998, \apjl, 492, L59

\bibitem[{{Tagger} \& {Pellat}(1999)}]{tagger99}
{Tagger}, M. \& {Pellat}, R. 1999, \aap, 349, 1003

\bibitem[{{Tagger} {et~al.}(2004){Tagger}, {Varni{\`e}re}, {Rodriguez}, \&
  {Pellat}}]{T04}
{Tagger}, M., {Varni{\`e}re}, P., {Rodriguez}, J., \& {Pellat}, R. 2004, \apj,
  607, 410

\bibitem[{{Titarchuk} \& {Fiorito}(2004)}]{titarchuk04}
{Titarchuk}, L. \& {Fiorito}, R. 2004, \apj, 612, 988

\bibitem[{{Varni{\`e}re} \& {Blackman}(2005)}]{varniere05}
{Varni{\`e}re}, P. \& {Blackman}, E.~G. 2005, \na, 11, 43

\bibitem[{{Varni{\`e}re} {et~al.}(2002){Varni{\`e}re}, {Rodriguez}, \&
  {Tagger}}]{varniere02}
{Varni{\`e}re}, P., {Rodriguez}, J., \& {Tagger}, M. 2002, \aap, 387, 497

\bibitem[{{Varni{\`e}re} \& {Tagger}(2002)}]{VT02}
{Varni{\`e}re}, P. \& {Tagger}, M. 2002, \aap, 394, 329

\bibitem[{{Varniere} {et~al.}(2011){Varniere}, {Tagger}, \& {Rodriguez}}]{V11}
{Varniere}, P., {Tagger}, M., \& {Rodriguez}, J. 2011, \aap, 525, A87

\bibitem[{{Varniere} {et~al.}(2012){Varniere}, {Tagger}, \& {Rodriguez}}]{V12}
{Varniere}, P., {Tagger}, M., \& {Rodriguez}, J. 2012, \aap, 545, A40

\bibitem[{{Veledina} {et~al.}(2013){Veledina}, {Poutanen}, \&
  {Ingram}}]{Veledina13}
{Veledina}, A., {Poutanen}, J., \& {Ingram}, A. 2013, \apj, 778, 165

\bibitem[{{Veledina} {et~al.}(2015){Veledina}, {Revnivtsev}, {Durant},
  {Gandhi}, \& {Poutanen}}]{Veledina15}
{Veledina}, A., {Revnivtsev}, M.~G., {Durant}, M., {Gandhi}, P., \& {Poutanen},
  J. 2015, \mnras, 454, 2855

\bibitem[{{Vincent} {et~al.}(2011){Vincent}, {Paumard}, {Gourgoulhon}, \&
  {Perrin}}]{Vin11}
{Vincent}, F.~H., {Paumard}, T., {Gourgoulhon}, E., \& {Perrin}, G. 2011,
  Classical and Quantum Gravity, 28, 225011

\end{thebibliography}
\bibliographystyle{aa}

\end{document}